\documentstyle[times,pramana,epsf,floats]{ias}
\newcommand{\beq}{\begin{equation}}
\newcommand{\eeq}{\end{equation}}
\newcommand{\be}{\begin{eqnarray}}
\newcommand{\ee}{\end{eqnarray}}
\def\nue{{\nu_e}}

\def\numu{{\nu_{\mu}}}

\def\nutau{{\nu_{\tau}}}

\newcommand{\dm}{\mbox{$\Delta{m}^{2}$~}}

\newcommand{\bee}{\mbox{$^{7}{Be}$~}}

\newcommand{\cnv}{\mbox{$\breve{\rm C}$erenkov~}}
\def\br{{$^{8}{B} ~$}}

\def\kl{{KamLAND~}}
\begin{document}

\title{Solar Neutrino Oscillation Phenomenology}

\author{Srubabati Goswami}
\address 
{Harish-Chandra Research Institute, Chhatnag Road, Jhusi,
Allahabad - 211-019, India}
\pacs{14.6q}
\abstract{This article summarises the status of the solar neutrino 
oscillation phenomenology  at the end of 2002 in the light of the
SNO and KamLAND results. We first present the allowed areas obtained from 
global solar analysis and demonstrate the  
preference of the solar data towards  the Large-Mixing-Angle (LMA) 
MSW solution.
A clear confirmation in favor of the LMA solution comes from the 
\kl reactor neutrino data. 
The  
\kl spectral data in conjunction with the global solar data   
further narrows down the allowed LMA region and splits it  
into two allowed zones -- a low $\Delta m^2$ region (low-LMA)
and 
high $\Delta m^2$ region (high-LMA). 
We demonstrate through a projected analysis that  
with an exposure of 3 kton-year (kTy) \kl can remove this ambiguity.
}

\maketitle
\section{The Neutrinos from the sun}
Solar neutrinos are produced via the reaction 
\beq
4p \rightarrow ^4He + 2e^+ + 2{\nue} + 28 {\rm MeV}
\eeq
The above process occurs through 
two main cycles of nuclear reactions -- the 
pp chain (CNO cycle) which is responsible for 98.5\% (1.5\%) 
of the energy. 
There are eight different types of neutrino fluxes, named according
to the 
parent nuclei of the decay chain which generates it.  
The pp chain gives rise to the neutrinos 
$pp,~~pep,~~hep,~~^7Be,~~^8B$ while the neutrinos 
$^{13}{N}$, $^{15}{O}$,$^{17}{F}$ 
are generated through nuclear reactions forming the CNO cycle. 
The solar neutrino fluxes are calculated by the so called
"Standard Solar Models" (SSM) among which the most extensively used are 
the ones due to Bahcall and his collaborators \cite{bp00}. 
The flux predictions from the SSM are robust. 
Different solar models agree to a very high degree of accuracy
( to within 10\%)
when the same input values of the parameters 
are used and also demonstrate striking 
consistency with helioseismological measurements.  
The $pp$ neutrinos are mainly responsible  
for solar
luminosity and the SSM prediction for the pp flux is least uncertain. 
The prediction for the $^8B$ neutrino  flux is most  uncertain 
stemming from the uncertainties associated with the cross-section 
of the reaction $^7{Be}(p \gamma)^8B$ producing these neutrinos. 

\section{Solar Neutrino Experiments}
The pioneering experiment for the detection of solar neutrinos is the 
$^{37}{Cl}$ experiment in Homestake which started operation in 1968 
{\footnote{
For recent reviews on solar neutrino experiments see 
\cite{Goswami:2003b,Miramonti:wz}.}}. 
It utilises the reaction \cite{cl}
\beq
\nu_e + ^{37}{Cl} \rightarrow ^{37}{Ar} + e^{-}.
\label{37cl}
\eeq
The threshold for this is  0.814 MeV and hence it is sensitive to
the $^{8}{B}$ and $^{7}{Be}$ neutrinos.    

Three experiments SAGE in Russia and GALLEX and its updated version GNO
in Gran-Sasso Underground laboratory in Italy uses the reaction 
\cite{ga}
\beq \nu _e \; + \; ^{71}Ga \; \rightarrow \;^{71}Ge\; +
\;e^- \label{gaeq} \eeq 
for detecting the solar neutrinos. 
This reaction has a low threshold of 0.233
MeV and the detectors are sensitive to the basic $pp$ neutrinos.

The radio chemical experiments $^{37}{Cl}$ and
$^{71}{Ga}$ experiments are sensitive to 
only $\nu_e$ and can provide the total solar $\nue$ flux. 

The first real time measurement of the solar neutrino flux was 
done by the Kamiokande imaging water \cnv detector, 
located in the Kamioka mine in Japan \cite{kam}.
It was subsequently upgraded to SuperKamiokande -- a same type of detector but
with much larger 
volume increasing the statistics \cite{superk}. 
The neutrinos interact with the electrons in the water via
\beq
e^{-} +  \nu_x \rightarrow e^{-} +  {\nu_x}
\label{nuescatt}
\eeq
This reaction is sensitive to all the three neutrino flavours.
However the $\numu$  and $\nutau$ react {\it via} the neutral
current which is suppressed by a factor of 1/6 compared to the
$\nu_e$ interaction which can be mediated by both charged and
neutral currents. 
The recoil electron energy threshold in Kamiokande was 7.5 MeV which could be 
reduced to 5 MeV in SuperKamiokande. 
Thus both the detectors are sensitive mainly to the \br neutrinos.
 
The Sudbury Neutrino Observatory (SNO) experiment also uses a \cnv detector 
but containing heavy water ($D_2O$). 
The deuterium in heavy water makes it possible to observe solar neutrinos 
in three different reaction channels \cite{snocc,snonc}
\begin{center} 
$\nu_e + d \rightarrow p + p +e^- $~~~~(CC) \label{nued} 
\end{center}
\begin{center} 
${\nu_x} + e^- \rightarrow {\nu_x} +e^- $~~~~(ES) 
\end{center} 
\begin{center}
$\nu_x + d \rightarrow n + p + \nu_x$~~~~~(NC) 
\label{neutral} 
\end{center}
The charged current (CC)  reaction is exclusive for $\nu_e$, 
The electron scattering (ES) reaction is 
same as in SK. 
The unique feature of SNO is the neutral current (NC) reaction 
which is sensitive to all the three flavours with equal strength. 
For both CC and ES reactions the final state electrons are directly detected 
through the \cnv 
light emitted by them which 
hits the PMTs and an event is recorded.
For the NC reactions the final state neutron can be captured 
(i) by another deuteron 
(ii)by capture on Cl in an NaCl enriched heavy water
(iii)by $^3{He}$ proportional counters.
For both (i) and (ii) the nuclei after capturing the neutrons emits 
single and multiple gamma rays respectively which compton scatters the 
electrons in the medium. 
The \cnv light produced by these electrons will produce an event. 
Therefore if the NC events are due to (i) and (ii) above then 
they cannot be disentangled from the CC and ES events 
Exclusive detection of the neutrons produced in the NC event
is possible for the process (iii). 

\subsection{The total solar neutrino flux}

The ratio of the observed solar neutrino rates to the
SSM predictions are presented below. 
 
\[
\begin{tabular}{ccc} \hline
experiment & $\frac{obsvd}{BPB00}$ \\ \hline {Cl} &
0.337 $\pm$ 0.029
\\
{Ga} & 0.553 $\pm$ 0.034
\\
{SK} & 0.465 $\pm$ 0.014
\\
{SNO(CC)} & 0.349 $\pm$ 0.021 
\\
{SNO(ES)} & 0.473 $\pm$ 0.074  
\\
{SNO(NC)} &  1.008 $\pm$ 0.123  
\\\hline
 \hline
\end{tabular}
\]
The declared  SNO NC data is due to neutron capture on deuteron. 
Therefore the CC,ES and NC events cannot be separated on an event
by event basis. 
For extracting the separate rates from the entangled data sample one 
one needs to assume  an  
undistorted \br flux as an input.  
The SNO rates quoted in the above Table are obtained under this assumption
\cite{snonc}. 
In all the above experiments 
observed $\nue$ flux is less than the theoretical predictions
implying disappearance of the solar $\nu_e$s. On the other hand for the 
SNO NC data the observed rate agrees to the theoretical prediction. 
Since the NC is sensitive to $\numu$ and $\nutau$ as well  this indicates that 
the $\nu_e$s are reappearing as $\numu$s and/or $\nutau$s.  

\subsection{Information on Direction and Energy}
Apart from providing a measurement for the total solar neutrino flux the 
real time measurements can also provide information on direction and 
energy of the incoming neutrinos. 
The electron scattering reaction used in SK and SNO has 
excellent directional sensitivity. 
In fact through this reaction the Kamiokande experiment first demonstrated 
the solar origin of the neutrinos. 
\begin{figure}
\epsfxsize=5cm
\vglue -1.0cm \hglue -3.8cm
\centerline{\epsfbox{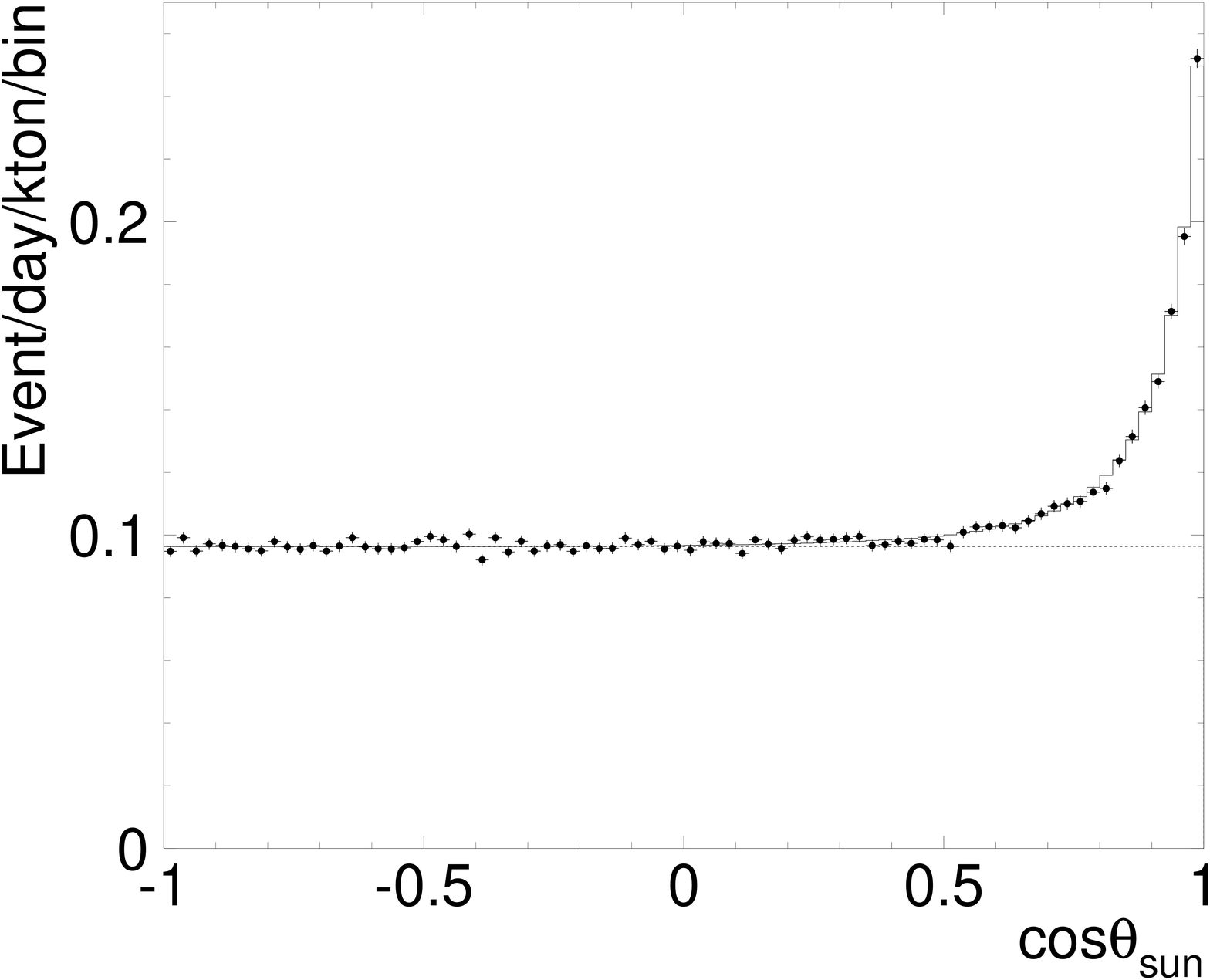}} 
\epsfxsize=5cm
\vglue -4.0cm \hglue 3.3cm
\centerline{\epsfbox{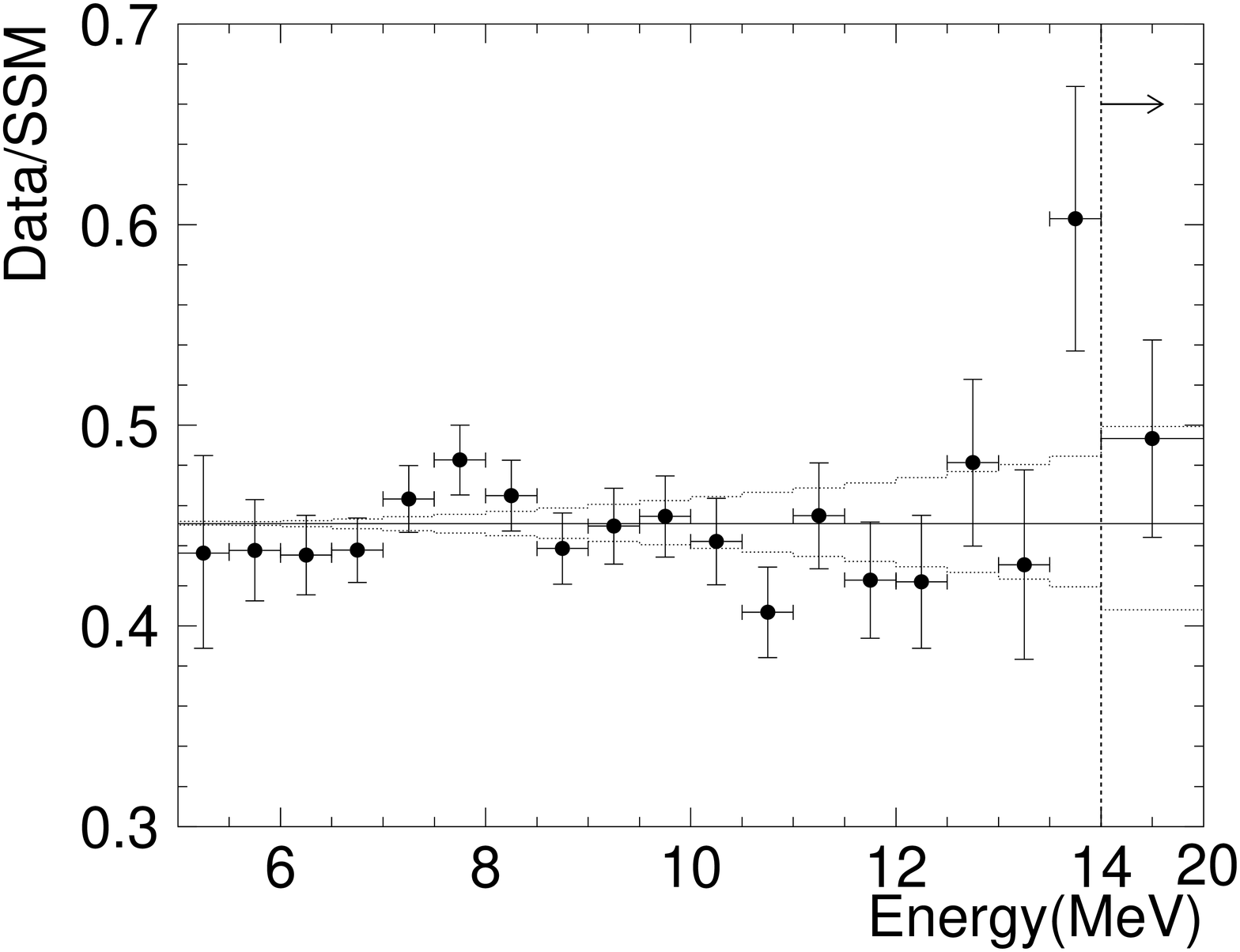}}
\caption{The left panel shows the number of events observed in SK 
as a function of $\cos\theta_{\odot}$ while the right panel shows the 
energy spectrum of the recoil electrons observed by SK.}
\label{skdir}
\end{figure}
The left panel in figure \ref{skdir} plots
the number of events  observed in SK against the 
cosine of the angle with the sun's direction. There is a clear peaking
towards $\cos\theta_{\odot}$ = 1.0 
\cite{Fukuda:2001nj}.
The statistical capacity of SK allows it to make an energy wise binning of the 
data and present the recoil electron energy spectrum.
The right panel in figure \ref{skdir} shows 
the data/SSM as a function of the recoil 
electron energy  with a 5 MeV threshold. 
The plot exhibits a flat recoil electron energy spectrum 
consistent with no spectral distortion 
\cite{Fukuda:2001nj}.

\begin{figure}[htbp]
\epsfxsize =4cm
\centerline{\epsfbox{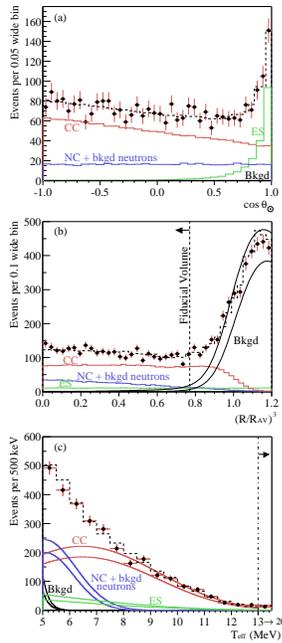}}  
\caption{The SNO spectrum vs $\cos\theta_{\odot}$, radial coordinate and 
recoil electron energy. Also shown is the MonteCarlo simulated events.}
\label{snospec}
\end{figure}

The panels 1 and 3 in 
Figure \ref{snospec} show the number of CC+ES+NC+background events in SNO 
as a function of direction and energy respectively. 
The solid lines in the  figures show 
the Montecarlo simulated CC,ES and NC events. 
The  ES events show the strong directional correlation
with sun as in SK. The CC events
has an angular correlation $1 - 0.34 cos
\theta_{\odot}$ while the NC events have no directional dependence as the 
produced gamma rays do not carry any information of the incident neutrino. 
The electron produced in the CC reaction being the only light
particle produced in the final state  has a strong correlation with the 
incident neutrino energy and thus can provide a good measurement of the 
$^8B$ energy spectrum. The recoil electron spectrum from the ES reaction is 
softer. 
The NC events due to capture on deuteron
has a peaked distribution around the energy of 6.25 MeV.  
These probability density functions are used to perform a maximum
likelihood fit to the data. 

\subsection{Variations with time}

SK already has enough statistics to divide their data into both energy 
and  zenith 
angle bins. The latest SK data has been presented 
with a binning of eighth energy bins and each energy bin contains data
subdivided into seven zenith angle bins. 
\cite{smy2002}.
This binning enables one to measure the  day time and night time fluxes
separately and study for any possible day/night asymmetry. 
Since in day time the neutrinos do not pass through the earths matter and
in night time they traverse the earths matter 
an observed  day/night asymmetry would be  
clear indicator of earths matter effect modifying the neutrino fluxes. 
The day/night asymmetry measured in SK  and SNO are still not at a 
statistically significant level. 

\subsection{Evolution of the solar neutrino problem}
Before the declaration of the SNO data there were two 
aspects of the
solar neutrino problem. 
The first one is -- in all the experiments 
the observed flux was less than  SSM prediction. 
The 
second one was the 'problem of the \bee neutrinos'. 
We note that among the pre-SNO experiments SK and its predecessor 
Kamiokande is  sensitive   to the \br neutrinos. 
The $^{37}{Cl}$ experiment is sensitive 
mainly to the \br and \bee neutrinos 
while the sensitivity of the $^{71}{Ga}$ experiments 
amount to $pp$, \br and \bee neutrinos. 
Combining 
the $^{8}{B}$ flux measured in SK with the Cl
data resulted in no room for the  $^{7}{Be}$
neutrinos. 
Similarly 
the expected $pp$ flux in Ga consistent with solar luminosity plus
the $^{8}{B}$ flux observed in SK indicated a negative   flux of 
$^{7}{Be}$ neutrinos in Ga.
The  solar physics could not explain this preferential vanishing of 
\bee flux over \br flux as in the pp chain \bee comes before \br 
and any mechanism that reduces the \bee flux would eventually 
reduce the \br flux. 
Therefore the answer was sought in the properties of neutrinos 
and neutrino flavour conversion was
considered as the most promising candidate for the solution. 
SNO provided the compelling evidence. 


The SNO CC reaction is sensitive only to $\nu_e$ while the 
ES reaction in both SK and SNO is sensitive  to both 
$\nu_e$ and $\nu_\mu/\nu_\tau$. 
Therefore a higher ES flux as compared to the CC flux 
will imply the presence of $\numu/\nutau$ in the solar $\nu_e$
flux. 
Combining the SK ES and SNO CC results one gets
\be
\phi_{ES}^{SK} - \phi_{CC}^{SNO} =
0.57 \pm 0.17 \times 10^{6} {\rm /cm^2/sec}
\ee
This is a 3.3$\sigma$ signal for $\nu_e$ transition to an active
flavour (or against
$\nu_e$ transition to solely a sterile state).

The  NC reaction is sensitive  to all the three flavours with equal strength 
resulting in a greater sensitivity   to the neutral 
current component in the solar $\nu_e$ flux. 
Comparing the NC and CC data from SNO 
one gets
\be
\phi_{NC}^{SNO} - \phi_{CC}^{SNO} = 3.41 \pm 0.65 \times 10^{6} 
{\rm /cm^2/sec}
\ee
In two circumstances we can have the CC and NC rates equal to each other.  
Either when there is no flavour conversion or for flavour conversion to a
 purely sterile state which does not interact with the detector. 
The observed CC/NC difference rules out both these possibilities 
at 5.3$\sigma$. 

\section{Two Flavour Oscillation}
If neutrinos have mass then the flavour eigenstates $\nue$ and $\numu/\nutau$
are different from the mass eigenstates $\nu_1$,$\nu_2$ and related as 
\be
\pmatrix{\nu_e \cr \nu_x}
= \pmatrix{ \cos\theta & \sin\theta
\cr -\sin\theta  & \cos\theta} \pmatrix{\nu_1 \cr \nu_2}
\ee
where $\theta$ is the mixing angle in vacuum. This leads to 
neutrino oscillation in vacuum \cite{bruno}.  
Then the survival probability that a $\nue$ remains   $\nue$
 after  traveling  distance L in vacuum is 
\be
P_{\nu_{e}\nu_{e}} =  1 - \sin{^2}2\theta \sin{^{2}}(1.27{\Delta
m}^{2}L/E)
\label{p2nu}
\ee
$\Delta m^2 = m_2^2 - m_1^2$. The    
term containing \dm is the oscillatory term 
resulting from  coherent propagation of the 
mass eigenstates. 

In matter, only $\nu_e$'s undergoes Charged current interaction
giving rise to an matter induced mass term of the form 
$\sqrt{2}G_FN_e$. 
This changes the mixing angles as 
\be
\tan2\theta_M
= \frac{\Delta m^2 \sin2\theta}{\Delta m^2 \cos2\theta -
2 \sqrt{2} G_F n_e E}
\label{thetam}
\ee
$n_e$ is the
electron density of the medium and $E$ is the neutrino energy. Eq.
(\ref{thetam}) demonstrates the resonant behavior of $\theta_M$.
Assuming $\Delta m^2  > 0$ the mixing angle in matter is maximal
(irrespective of the value of mixing angle in vacuum) for an
electron density satisfying,
\begin{equation}
2\sqrt{2}G_{F}n_{e,res}E = \Delta m^2 \cos 2\theta \label{neres}
\end{equation}
This is
the Mikheyev-Smirnov-Wolfenstein (MSW) effect 
of matter-enhanced
resonant flavor conversion \cite{msw}.

The most general expression for $\nu_e$ survival probability in an
unified formalism  over the mass range $10^{-12} - 10^{-3}$ eV$^2$
and for the mixing angle $\theta$ in the range [0,$\pi/2$] is
\cite{petcov} 
\be P_{ee}&=&P_{\odot}P_{\oplus} + (1-P_{\odot})
(1-P_{\oplus}) \nonumber \\ && + 2\sqrt{P_{\odot}(1-P_{\odot})
P_{\oplus}(1-P_{\oplus})}\cos\xi 
\label{probtot} 
\ee 
where
$P_{\odot}$ denotes the probability of conversion of $\nu_e$ to
one of the mass eigenstates in the sun and $P_{\oplus}$ gives the
conversion probability of the mass eigenstate back to the $\nu_e$
state in the earth. All the phases involved in the Sun, vacuum and
inside Earth are included in $\xi$.
Depending on the value of $\Delta m^2/E$ one has the following three limits 
\\
(i)in the regime $\Delta m^2/E \stackrel{<}{\sim}
5\times 10^{-10}$ eV$^2$/MeV matter effects inside the Sun
suppress flavor transitions while the effect of the phase $\xi$ 
remains. This is the vacuum oscillation limit.
\\ 
(ii)For $\Delta m^2/E \stackrel{>}{\sim} 10^{-8}$
eV$^2$/MeV, the total oscillation phase
becomes very large and the $\cos\xi$ term in Eq. (\ref{probtot})
averages out to zero signifying incoherent propagation of 
the neutrino mass eigenstates. 
This is the MSW limit. 
\\
(iii)For 
$5\times 10^{-10}$ eV$^2$/MeV $\stackrel{<}{\sim}
\Delta m^2/E \stackrel{>}{\sim} 10^{-8}$ eV$^2$/MeV, both matter effects
inside the Sun and coherent oscillation effects in the
vacuum become important. This is the {\it quasi vacuum oscillation}
(QVO) regime.

\begin{figure}[htbp]
\epsfxsize=6cm
\vglue -0.5cm \hglue -3.8cm
\centerline{\epsfbox{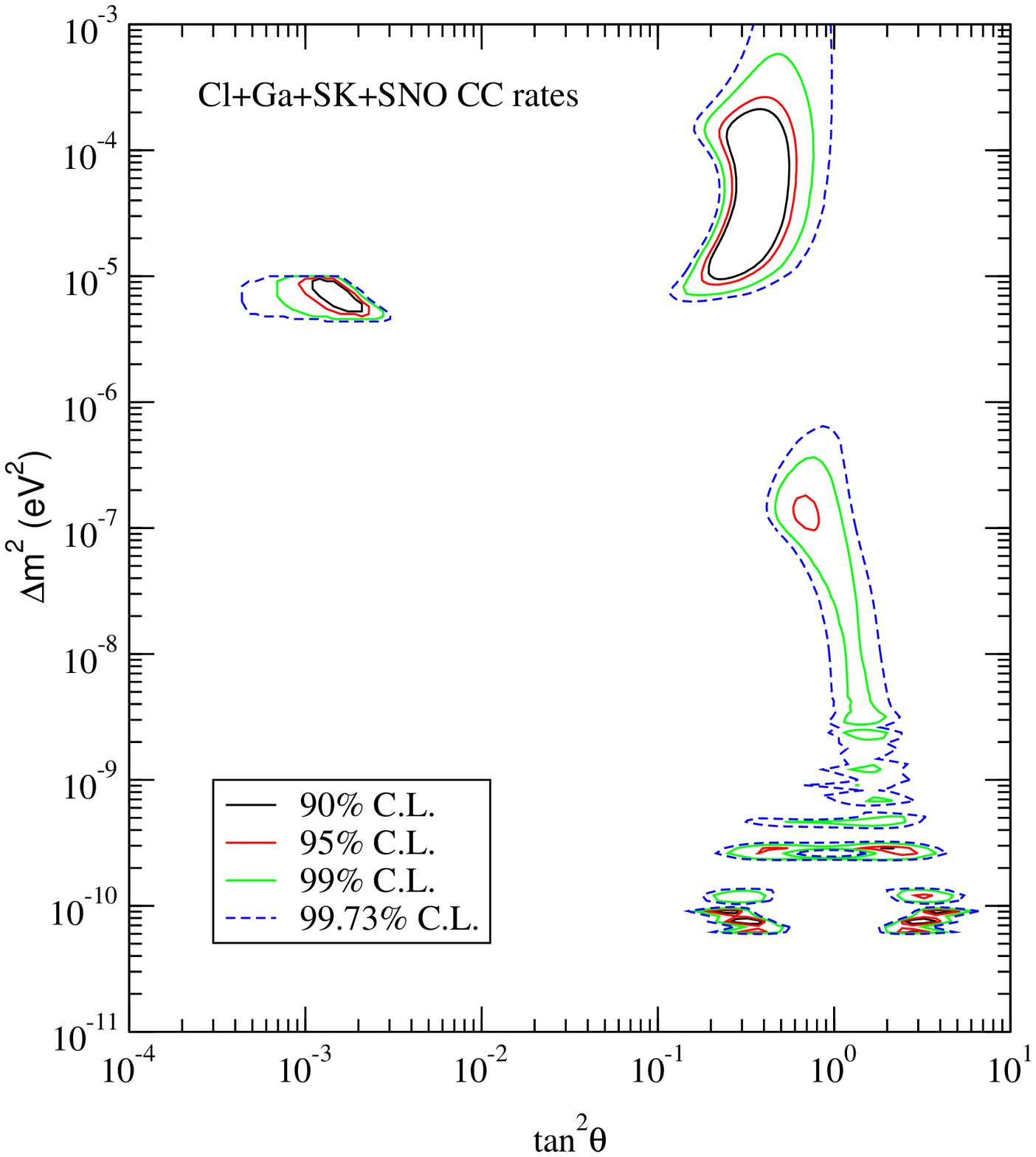}} 
\epsfxsize=6cm
\vglue -8.5cm \hglue 3.3cm
\centerline{\epsfbox{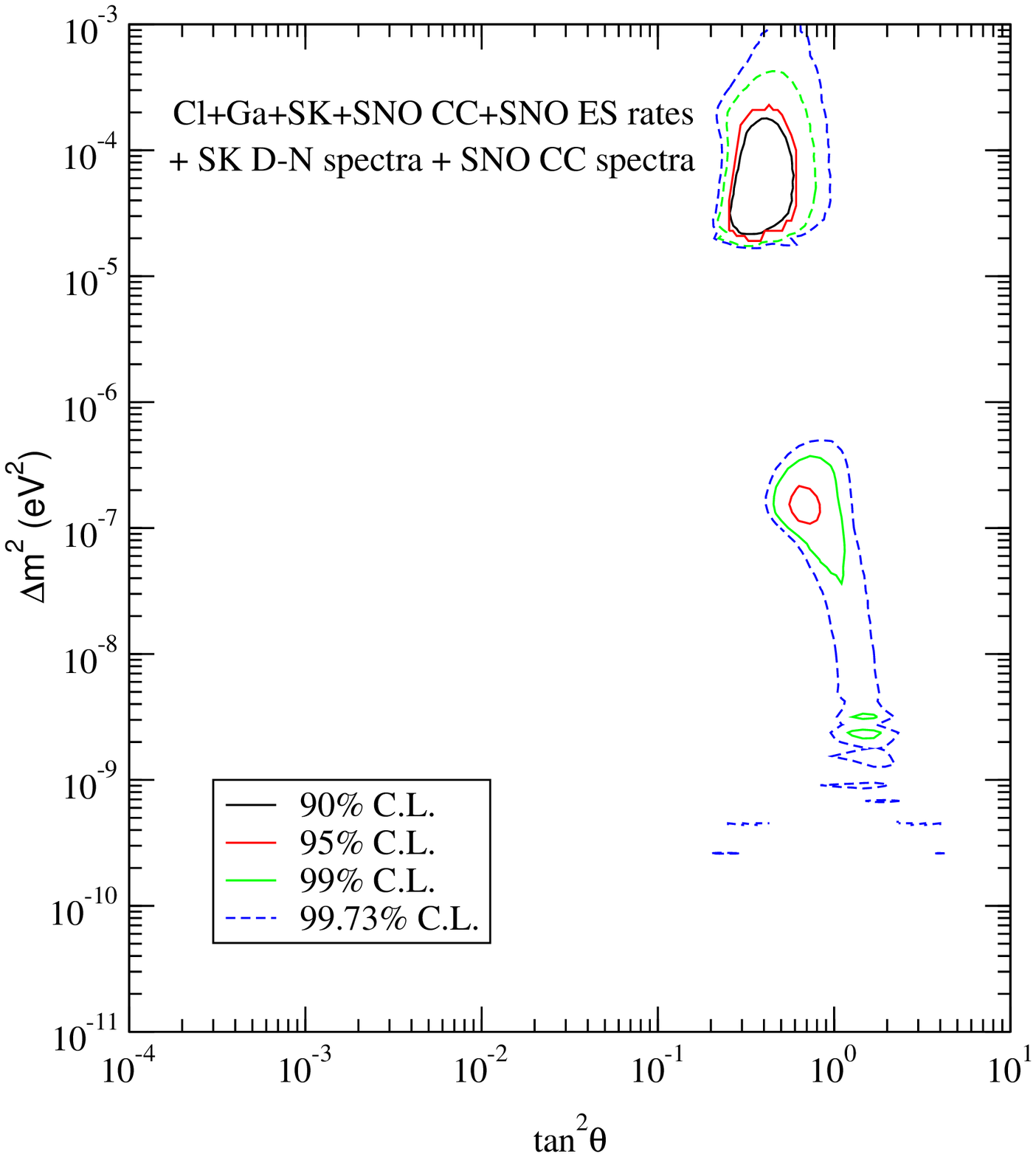}}
\caption{The allowed regions in \dm-$\tan^2\theta$ plane from data on total 
solar rates in the left panel and from data on total solar rates and SK 
spectrum data in right panel.}
\label{snocc}
\end{figure}
Next we present the results obtained by performing 
a $\chi^2$-analysis of the solar neutrino data. 
The procedure followed can be found in  
\cite{Choubey:2001bi,Choubey:2001ws,our-snocc}.
In figure \ref{snocc} we present the allowed regions obtained from 
analysis including the total fluxes measured in Cl, Ga, SK and SNO  
\cite{our-snocc}. 
There are basically five regions which are allowed --  
 the small mixing angle region (SMA), 
the large mixing angle high \dm regions (LMA),  
the large mixing angle-low \dm regions (LOW),
the vacuum oscillation regions 
symmetric about $\tan^2\theta=1.0$ and the 
quasi vacuum region between the LOW and the vacuum oscillation regions. 

In the scond panel of figure \ref{snocc}
 we present the allowed areas after including the 
SK spectrum data with the total rates data \cite{our-snocc}. 
The SMA region and large part of the vacuum  oscillation region 
are seen to have been washed away with the inclusion of the SK spectrum data. 

\begin{figure}[htbp]
\epsfxsize=6cm
\vglue -0.5cm \hglue -4.2cm
\centerline{\epsfbox{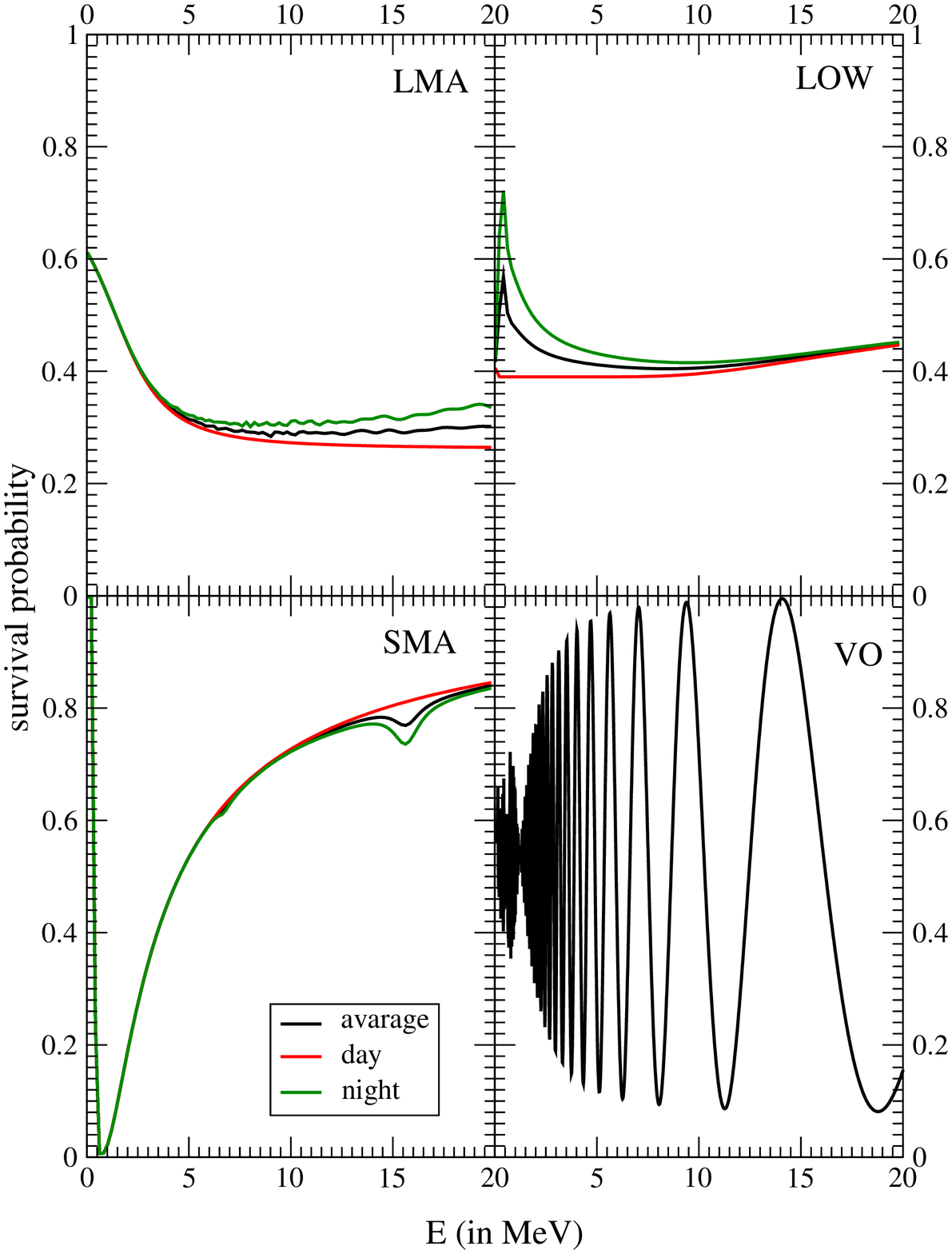}}
\epsfxsize=6cm
\vglue -7.6cm \hglue 3.3cm
\centerline{\epsfbox{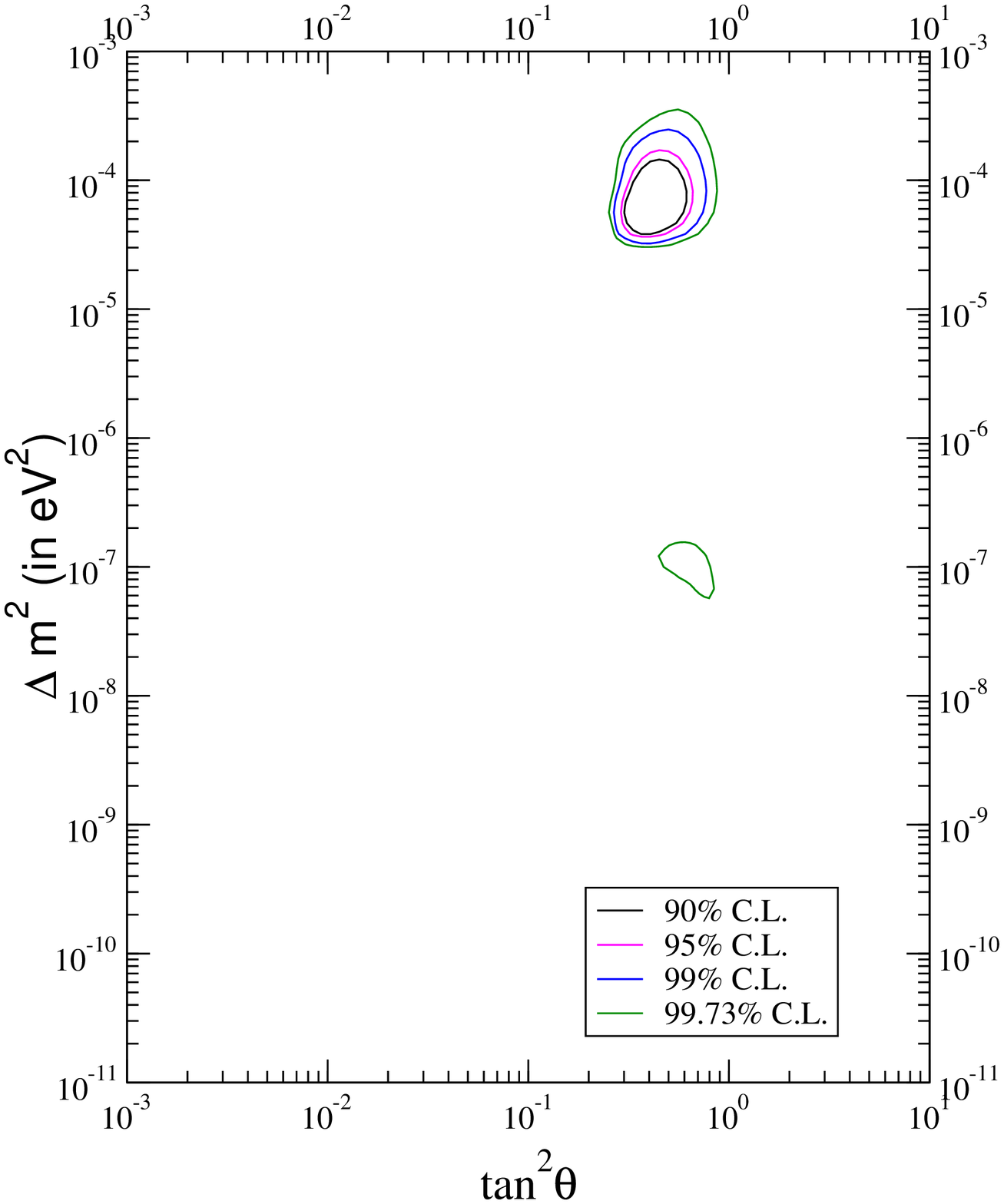}}
\caption{The left panel shows the dependence of the probability on energy.
The right panel gives the allowed area after including the latest SNO 
spectrum data including the NC events. } 
\label{prob}
\end{figure}

In the left panel of 
fig. \ref{prob} we show the dependence of the probabilities on energy. 
In the SMA and the VO oscillation  regions the probability has a 
non-monotonic dependence on 
energy whereas in the LMA and LOW regions  the survival probability 
does not have any appreciable dependence on energy beyond 
5 MeV which is the threshold for SK and SNO. 
Thus these regions are favoured by the 
flat SK spectrum.  

In the right panel of  figure \ref{prob} ( from \cite{Choubey:2002nc}) 
we present the allowed regions
obtained from 
an global $\chi^2$ analysis 
\cite{Choubey:2002nc,our-snonc2,Bandyopadhyay:2002qg,others}
 of the total rates observed in Cl and
Ga (SAGE, GALLEX and GNO combined rate), the 1496 day SK
zenith angle energy spectrum data and the recent SNO data of  
combined
CC+ES+NC+background  data in 17 day and 17 night energy bin.
After the inclusion of the SNO results
\\
(i) LMA emerges as the 
favoured solution.
\\
(ii) The LOW region now appears only at 3$\sigma$. 
\\
(iii)Values  of $\Delta m^2$ above $3\times 10^{-4}$
eV$^2$  are seen to be disfavored.
\\
(iv)The QVO and  VO solution are 
not allowed at $3\sigma$.\\
(v) Maximal mixing 
($\theta=\pi/4$) is disfavored at $3.4\sigma$.\\
(vi) 
SMA solution is disfavored at 3.7$\sigma$\\
(vii)The Dark Side ($\theta > \pi/4$)
solutions disappear. 

Apart from including the SNO spectral data the  figure \ref{prob} 
also has the new 1496 day SK zenith angle spectrum data. 
The data reveal a flat zenith angle spectrum. 
The inclusion of this data rule out the 
higher part of the LOW solution beyond $1.5\times 10^{-7}$ eV$^2$ 
for which peaks in the zenith-angle spectrum are expected
\cite{Gonzalez-Garcia:2000dj}. 

\section{KamLAND} 

\begin{figure}[htbp]
\epsfxsize=6 cm
\vglue -0.2cm \hglue -4.0cm
\centerline{\epsfbox{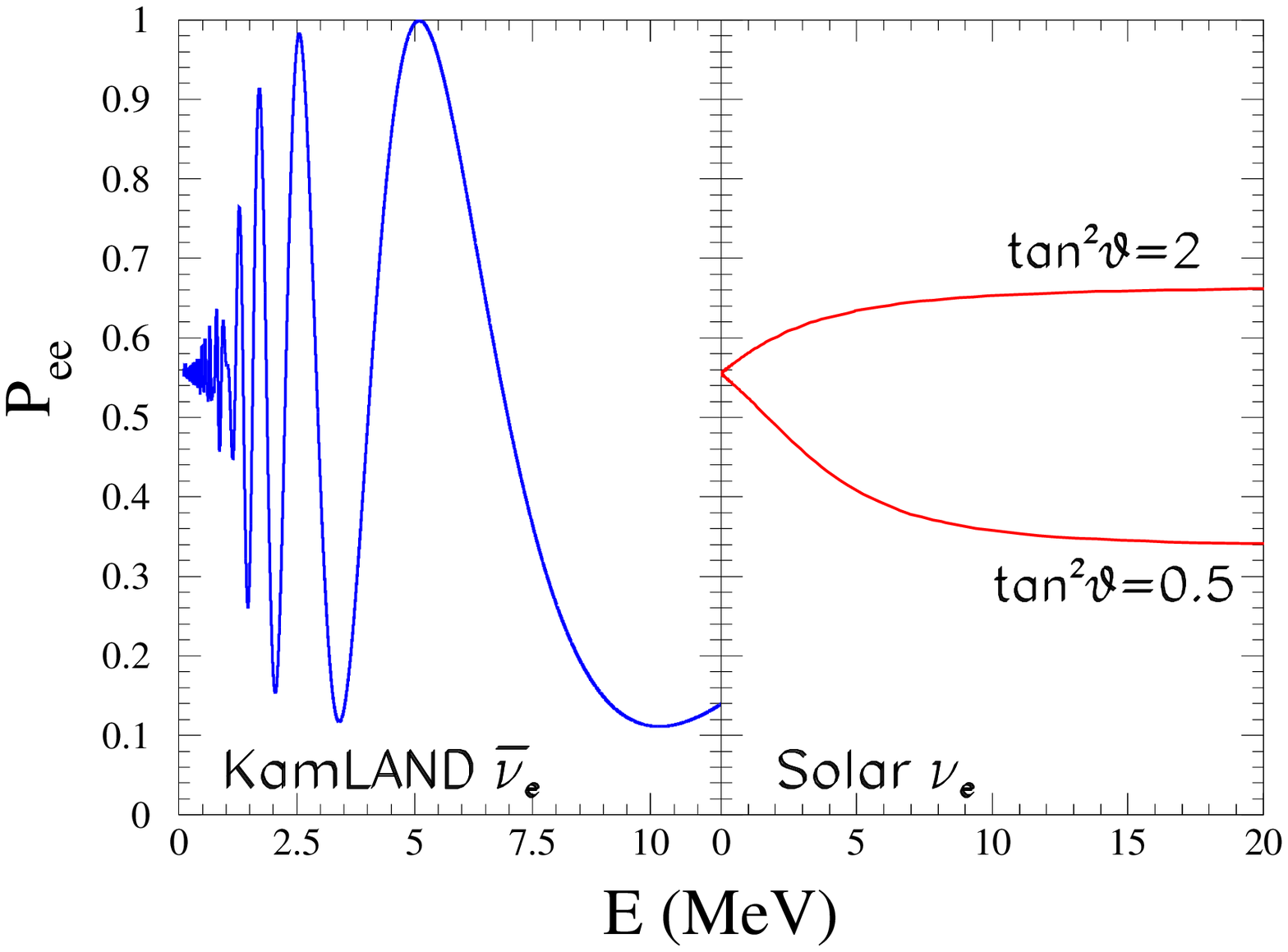}}
\epsfxsize=3.0cm
\vglue -4.4cm \hglue 3.3cm
\centerline{\epsfbox{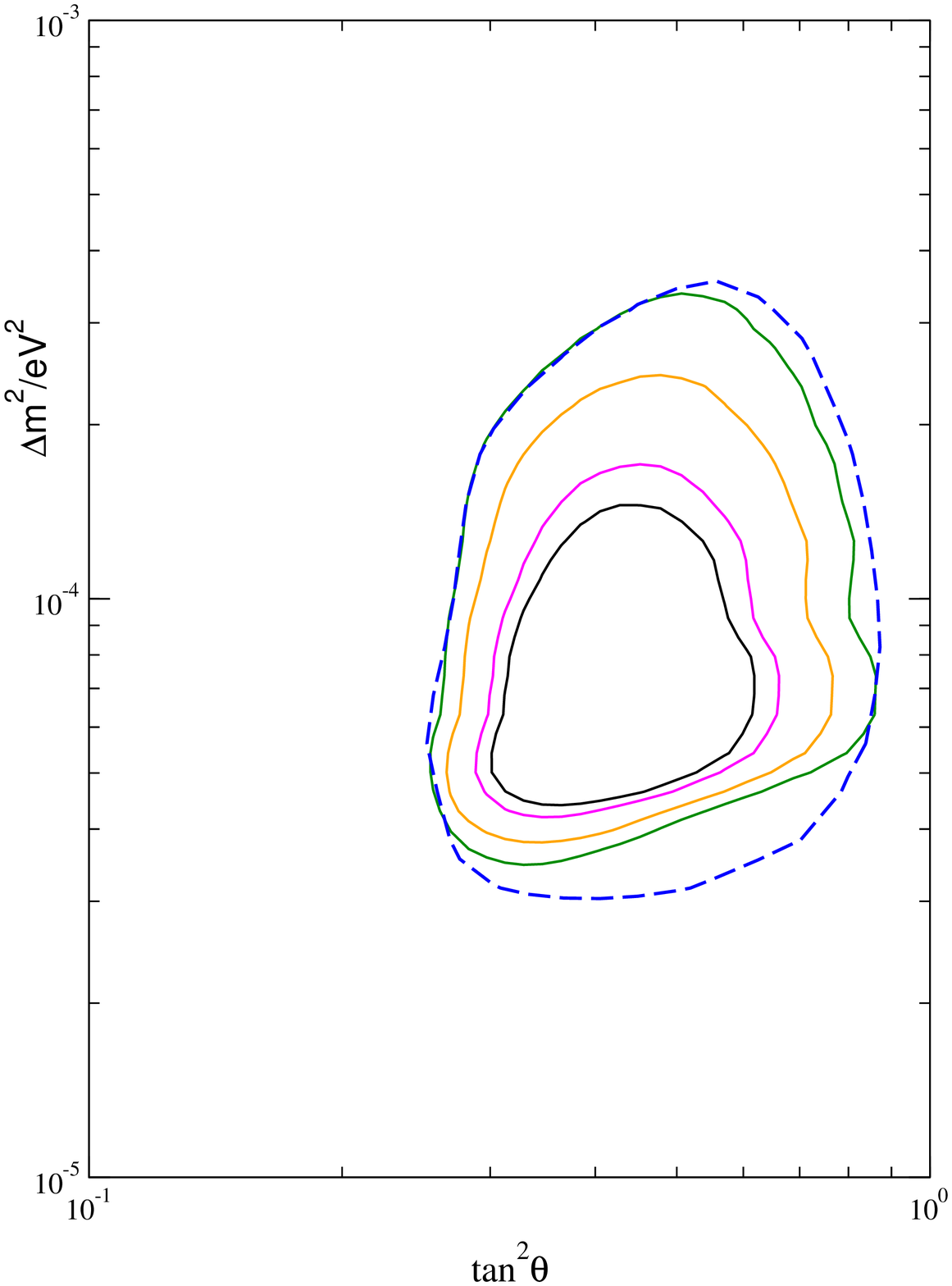}}
\caption{The left figure shows  the variation of the survival probability with 
energy for \kl and for the solar neutrinos. 
The right figure gives the allowed region from an analysis of \kl   
rate and global solar data.
The allowed region from only solar analysis is shown also by 
the outer dashed line.}
\label{klrate}
\end{figure}

KamLAND is a 
1000 ton liquid scintillator neutrino detector
situated at the former site of Kamiokande \cite{kl}. 
It looks for ${\bar{\nu_e}}$ oscillation coming from 16
reactors
at distances 81 - 824 km. 
Most powerful reactors are at a distance $\sim$ 180 km. 
The detection process is $\bar{\nu_e}p \rightarrow e^{+} n$. 
The $e^+$ produced annihilates  with the electrons in the medium to produce 
prompt photons. The neutrons get absorbed by the protons in the medium to 
produce 
delayed photons. The correlation of time,position and energy between these two 
constitute a grossly background free signal. 
The survival probability  relevant in KamLAND  
is the  vacuum  survival probability  (cf. eq. \ref{p2nu})
summed  over the
distances from all the reactors. 
The average energy and length scales for KamLAND are 
$E_{\nu} \sim $ 3 MeV, L $\sim$ 1.8 $\times 10^{5}$ m which makes it
sensitive to $\Delta m^2
\sim 1.6 \times 10^{-5}$ eV$^2$ which is in the LMA region. 
The figure \ref{klrate} shows the probabilities   for KamLAND  
for the  average distance of 180 km 
and solar neutrinos for  \dm (=$ 6\times10^{-5}$ eV$^2$) 
and $\tan^2\theta$ (=0.5) \cite{Bahcall:2002ij}. 
Whereas for the solar probabilities the \dm dependence is completely averaged 
out in KamLAND 
the probability exhibits a $L/E$ dependence which gives 
it an unprecedented sensitivity   to determine \dm.

The solar best-fit predicts a rate in \kl  
$0.65^{+0.08}_{-0.39}$ (3$\sigma$) \cite{raj}
while the observed rate is 
$0.611 \pm 0.094$  corresponding  to 145 days of data. 
Thus KamLAND confirms the LMA solution. 
In figure \ref{klrate} we show the allowed area obtained from 
a $\chi^2$-analysis of KamLAND rates data and global solar data
\cite{Bandyopadhyay:2002en}.  
The contour obtained from only solar analysis is also drawn (dashed lines) 
for understanding the role played by \kl. 
The inclusion of the \kl rates data gives a lower bound 
of  $\Delta m^2 > 4 \times 10^{-5}$ eV$^2$. 
The other parts of the parameter space allowed from the solar analysis 
are not constrained much. 

\begin{figure}[htbp]
\epsfxsize=5.0cm
\vglue -0.0cm \hglue -4.0cm
\centerline{\epsfbox{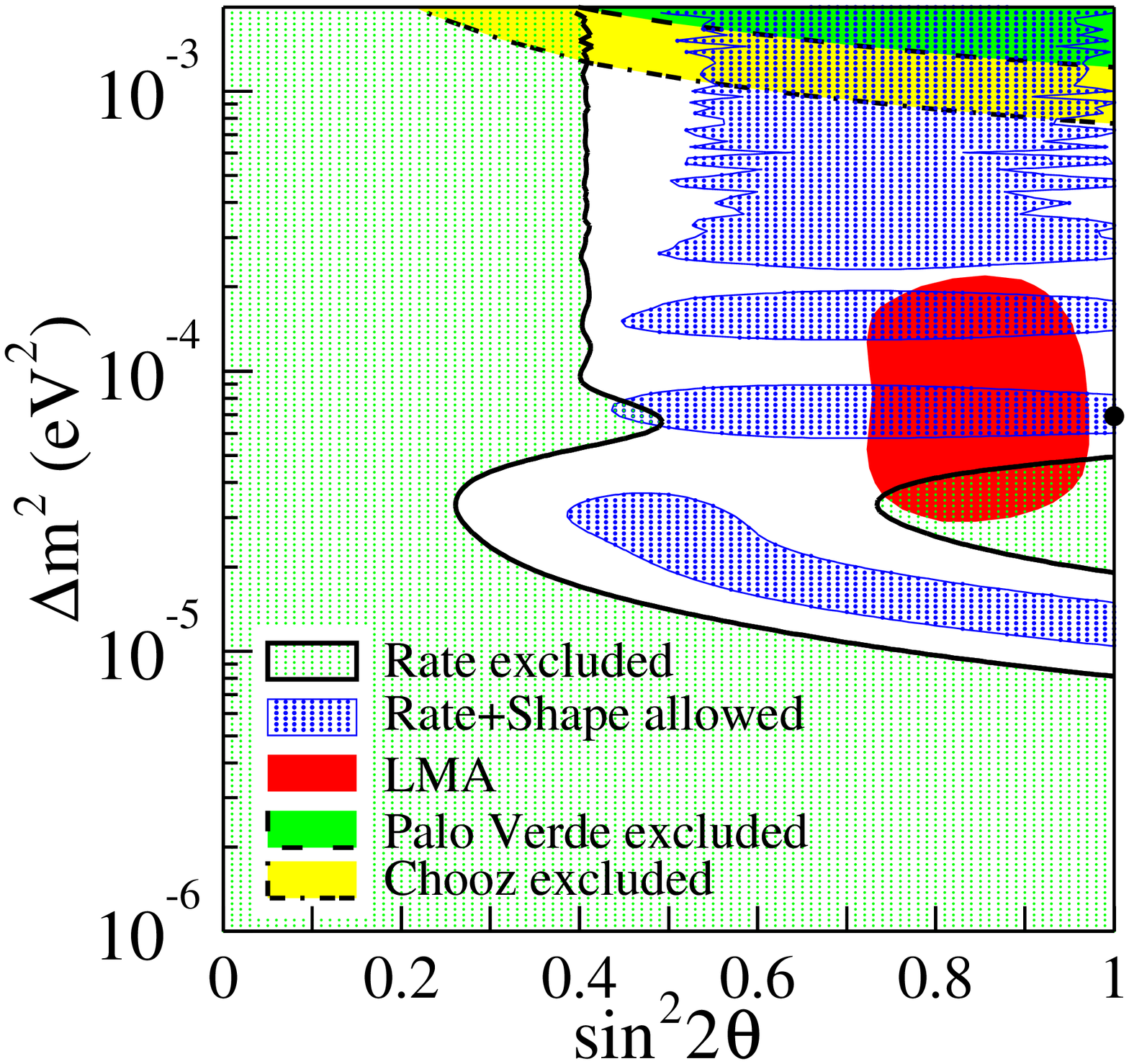}}
\epsfxsize=3.8cm
\vglue -5.2cm \hglue 3.3cm
\centerline{\epsfbox{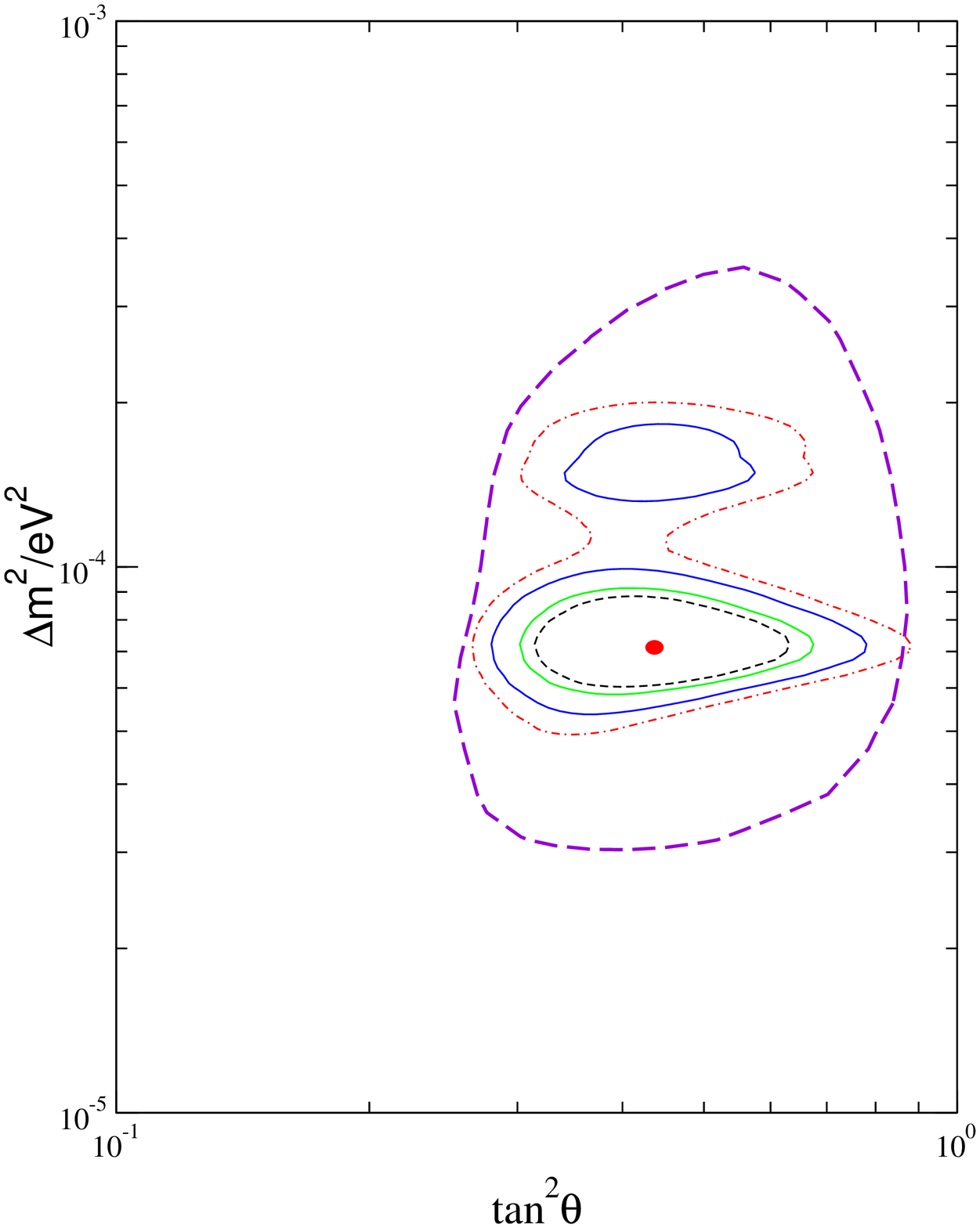}}
\caption{The left panel shows the allowed regions in the $\dm-\tan^2\theta$ 
plane from analysis of \kl spectrum data. 
The right panel shows the allowed region from a global 
analysis of \kl spectrum and 
solar neutrino data.}
\label{all}
\end{figure}

Figure \ref{all} 
shows the allowed areas obtained from a $\chi^2$ analysis of the 
only \kl spectrum \cite{kl}. 
For an average energy of 5 MeV and a distance 180 km a \dm of 
$7 \times 10^-5 $ eV$^2$ corresponds to the oscillation wavelength
($\lambda$)  
$\sim$ the distance traveled(L) and $P_{ee} \approx 1$, for \dm 
of $3.5 \times 10^{-5}$ eV$^2$ $\lambda=2L$ and $P_{ee} = 0.0$. Again for 
\dm = $1.4 \times 10^{-4}$ eV$^2$ $\lambda = L/2$ and $P_{ee} =1$. 
Since the \kl spectral data
 corresponds to a peak around 5 MeV islands around the
first and the third \dm are allowed whereas the middle \dm is 
disfavoured as is seen from the panel 1 of figure \ref{all}.   

In the second panel of figure \ref{all}
we show the allowed area from \kl spectrum and
global solar data  obtained  through a $\chi^2$ analysis 
\cite{Bandyopadhyay:2002en}.   
Inclusion of the \kl spectral data 
splits the allowed region into two zones at 99\% C.L. 
low-LMA (LMA1) and high-LMA (LMA2).  LMA2 has less statistical significance 
(by
$\approx 2\sigma$)
The global best-fit comes at $\Delta m^2 =7.17
\times 10^{-5}$ eV$^2$ and 
$\tan^2\theta = 0.44$ in low-LMA.
LOW region is disfavoured at $5\sigma$.
Maximal mixing although allowed by the \kl spectrum data gets disfavoured 
at 3.4$\sigma$ by the overall analysis. 

\begin{figure}[htbp]
\epsfxsize=6 cm
\vglue -0.4cm \hglue -3.8cm
\centerline{\epsfbox{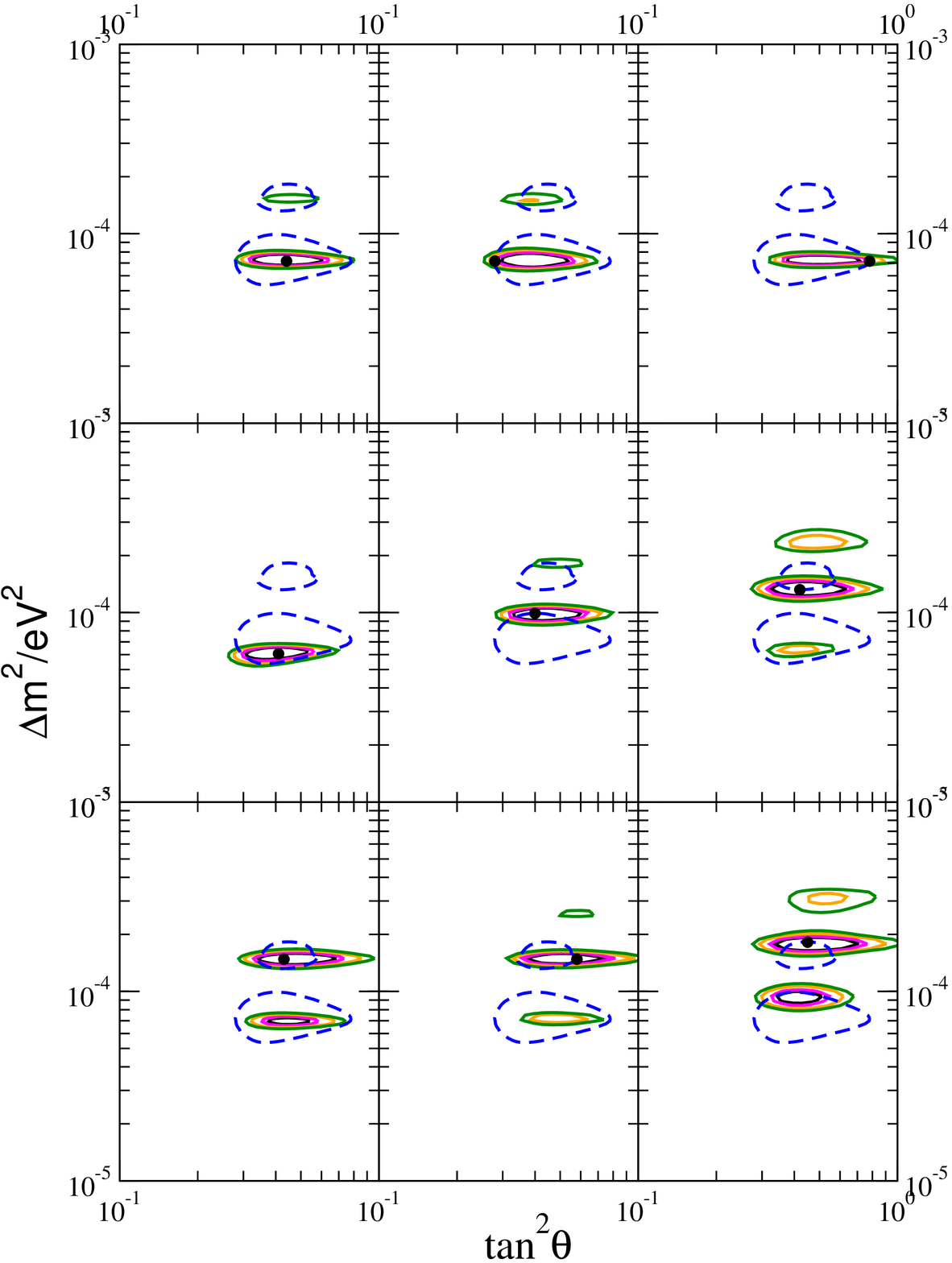}}
\vglue -8.5cm \hglue 3.3cm
\epsfxsize=6 cm
\centerline{\epsfbox{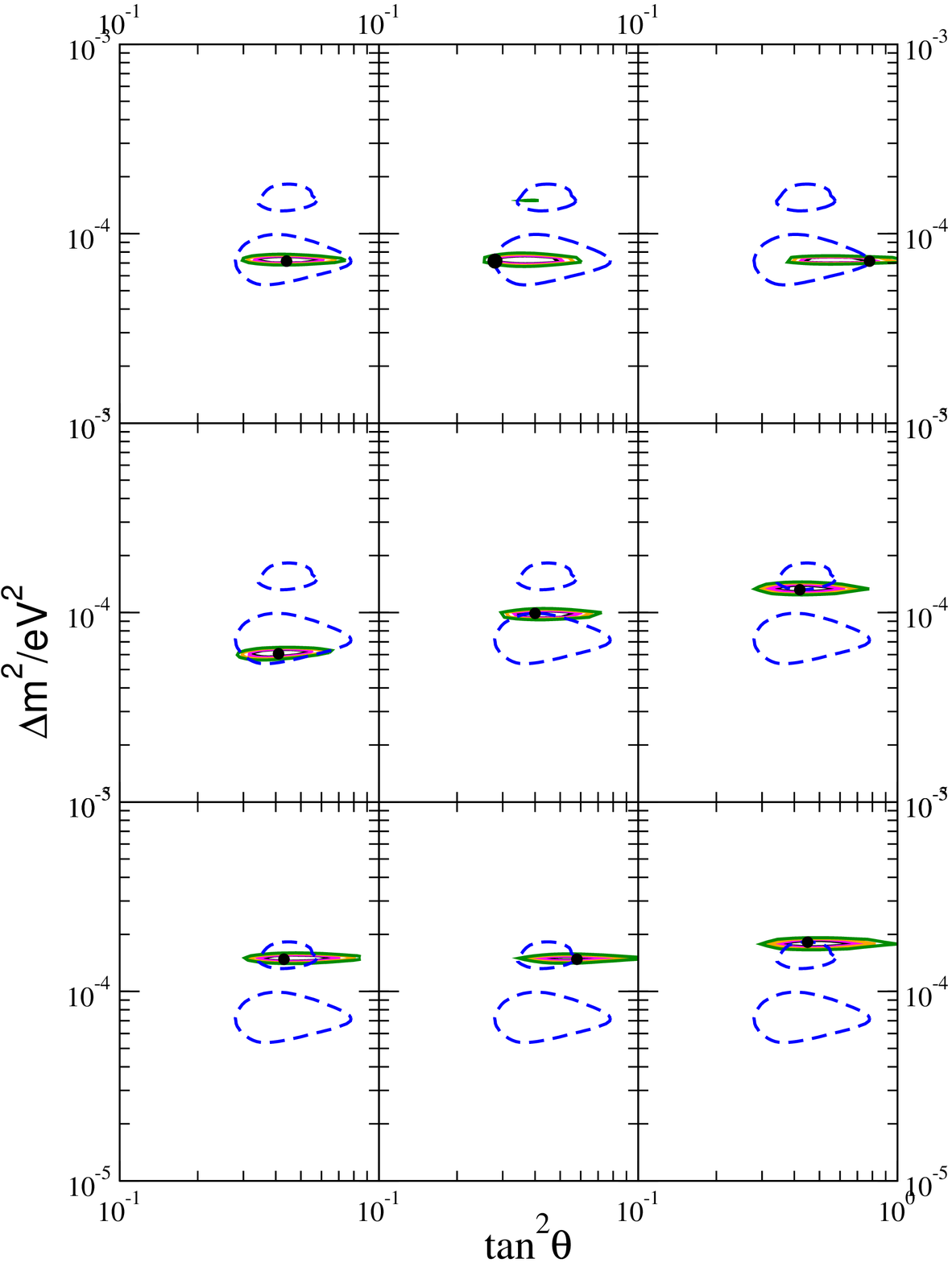}}
\caption{The 1 kTy (left panel) and 3 kTy (right panel)
projected contours of KamLAND and solar data. The points at which the spectra
are simulated are shown by bold dots.}
\label{proj}
\end{figure}

In the left panel of figure \ref{proj}
we explore through a projected analysis of 
1 kton year simulated spectrum the potential of \kl in 
distinguishing between the two allowed areas. 

The allowed $\Delta m^2$
ranges around both
low-LMA and high-LMA zones decrease in size. 
Since the solar data favours the low-LMA zone the
allowed areas become more precise for spectrum simulated at the low-LMA
zone while ambiguity
between the two zones remains for high-LMA  spectrum.
In the right panel of \ref{proj} we show that a higher statistics (3 kton year)
from \kl can remove
this ambiguity almost completely determining 
\dm to within 6\%. 

Analysis of solar and \kl data has been done in the realistic
three neutrino scenario in 
\cite{Fogli:2002au}. 
In this case the non-zero value of the mixing angle 
$\theta_{13}$ \footnote{this is at present limited by the CHOOZ data 
to  $\sin^2\theta_{13} < 0.04$.} 
connecting the atmospheric sector and the solar 
sector modifies the
allowed area in $\Delta m^2 - \tan^2\theta$
parameter space. 
A third solution at a \dm higher than the high-LMA zone
gets marginally allowed for the three flavour case. 
As $\theta_{13}$ is increased the high-LMA zones tend to 
disappear. 

\section{Conclusions}
The solar neutrino research began 
with a motivation of
understanding the sun through the neutrino channel. Over the years it
metamorphosed
into a tool of unraveling the fundamental properties of the neutrinos.
At the end of 2002 the status of the solar neutrino oscillation
phenomenology can be summarised as 
\\
{$\bullet$}
Comparison of SNO CC and SNO NC signifies neutrino
flavour conversion at $5.3\sigma$.
\\
{$\bullet$} Rules out transitions to
pure sterile  states at $5.3\sigma$.
\\
$\bullet$ {LMA} is the favoured solution to the solar
neutrino problem with $3\times 10^{-5}$ eV$^2 \leq \Delta m^2 \leq 3\times
10^{-4}$ eV$^2$ and $0.25 \leq \tan^2\theta \leq 0.87$ at $3\sigma$.
\\
$\bullet$
KamLAND confirms LMA.
\\ 
{$\bullet$} Best-fit $\Delta m^2$ shifts from  
$6 \times 10^{-5}$ eV$^2$ as obtained from global solar analysis to 
$7.2 \times 10^{-5}$ eV$^2$.\\
{$\bullet$}
There is no significant change in best-fit $\theta$ {($\tan^2\theta =0.4$)}.
\\
{$\bullet$}{LMA region splits into two parts at 99\% C.L.}.
\\
{$\bullet$}{$3\sigma$ allowed range after the first KamLAND data
is {$4.96 \times 10^{-5}$ eV$^2 < \Delta m^2 < 2 \times 10^{-4}$ eV$^2$} and
{$0.27 < \tan^2\theta < 0.88$}}.
\\
{$\bullet$}Transitions to a mixed state is still allowed with
$<13\%(52\%)$ sterile mixture at $1\sigma(3\sigma)$
\cite{Bahcall:2002ij}.
\\
At this juncture  
the emerging goals in solar neutrino research are
\\
(i)precise determination of the oscillation parameters;
\\
(ii)to observe the low energy end of the solar neutrino spectrum
consisting of the pp,CNO and the $^7{Be}$ line and do a full
solar neutrino spectroscopy.\\
As far as precision determination of \dm is concerned \kl has remarkable 
sensitivity and 3 kton year \kl data can remove the ambiguity between the 
two presently allowed zones completely. 
Even before that a day-night asymmetry $>1$\%
in SNO can exclude LMA2 \cite{deHolanda:2002iv}. 
However the constraining power of \kl for $\theta$ is not as good 
being limited by the 6\% systematic error and also by the fact that 
in the statistically significant regions 
the observed \kl spectrum  corresponds to a peak in the survival probability
where the $\theta$ sensitivity is
very low.  
At the present value of the best-fit parameters in the low-LMA region
a new \kl like reactor experiment
with a baseline of $\sim$ 70 km will be sensitive to the 
the minimum in the survival probability
increasing the $\theta$ sensitivity by a large amount 
\cite{Bandyopadhyay:2003du}.   
The upcoming 
Borexino experiment should see a rate {$0.64 \pm 0.03~(1\sigma)$ }
and no day-night asymmetry.
It cannot differentiate between the two LMA regions.
However it can provide a measurement of the $^7{Be}$ flux coming from the sun.
Real time measurement of the $pp$ neutrino flux is the target of the so 
called "LowNU" experiments like 
XMASS, HELLAZ, HERON. CLEAN, MUNU, GENIUS using $\nu-e$ scattering 
and  LENS, MOON and SIREN using charged current reactions \cite{scho}. 
All these experiments involve development of new and challenging 
experimental concepts and research work is in progress 
for evolving these techniques. 

I acknowledge my collaborators A. Bandyopadhyay, S. Choubey, R. Gandhi and 
D.P. Roy.


\begin{thebibliography}{99}
\bibitem{bp00} J. N. Bahcall, M.H. Pinsonneault and S. Basu,
Astrophys. J. {\bf 555} (2001)990.

\bibitem{Goswami:2003b}
S.~Goswami,
arXiv:hep-ph/0303075

\bibitem{Miramonti:wz}
L.~Miramonti and F.~Reseghetti,
Riv.\ Nuovo Cim.\  {\bf 25N7} (2002) 1
[arXiv:hep-ex/0302035].

\bibitem{cl} B. T. Cleveland  {\it et al.,} Astroph. J. {\bf 496}
(1998) 505.

\bibitem{ga}
J.N. Abdurashitov et al., (The SAGE collaboration),
astro-ph/0204245;  W. Hampel {\em et al.}, (The Gallex collaboration),
Phys. Lett. {bf B447}, 127 (1999);
M. Altmann {\it et al.}, (The GNO collaboration),Phys. Lett. {\bf
B492},16 (2000).

\bibitem{kam}Y. Fukuda {\em et al.}, (The
Kamiokande collaboration), {Phys. Rev. Lett.} {\bf 77}, 1683
(1996).

\bibitem{superk}
S.~Fukuda {\it et al.}  [Super-Kamiokande Collaboration],
Phys.\ Lett.\ B {\bf 539}, 179 (2002). 

\bibitem{snocc}Q.R. Ahmad {\it et al.},
(The SNO Collaboration)
Phys. Rev. Lett. {\bf 87}, 071301 (2001). 

\bibitem{snonc}
Q.~R.~Ahmad {\it et al.}  (SNO Collaboration),Phys.\ Rev.\ Lett.\  {\bf 89}, 011301 (2002);
Phys.\ Rev.\ Lett.\  {\bf 89}, 011302 (2002). 

\bibitem{Fukuda:2001nj}
S.~Fukuda {\it et al.}  [Super-Kamiokande Collaboration],
Phys.\ Rev.\ Lett.\  {\bf 86} (2001) 5651.

\bibitem{smy2002}
M.~B.~Smy, hep-ex/0202020.

\bibitem{bruno}B. Pontecorvo, JETP {\bf{6}}, 429 (1958); 
Z. Maki, M.
Nakagawa and S. Sakata, Prog. Theor. Phys. {\bf{28}}, 870 (1962).

\bibitem{msw}
L. Wolfenstein, {\em Phys. Rev.} {\bf D34}, 969 (1986);
S.P. Mikheyev and A.Yu. Smirnov,  {\em Sov. J. Nucl.
Phys.} {\bf 42(6)}, 913 (1985); {\em Nuovo Cimento} {\bf 9c}, 17 (1986).

\bibitem{petcov}S.T. Petcov,  Phys.
Lett. B214 (1988) 139; Phys. Lett. B406, 355 (1997);
G.L. Fogli, E.Lisi, D. Montanino and A. Palazzo,
Phys. Rev. {\bf D62}, 113004, (2000).

\bibitem{Choubey:2001bi}
S.~Choubey, S.~Goswami and D.~P.~Roy,
Phys.\ Rev.\ D {\bf 65}, 073001 (2002)
S.~Choubey, S.~Goswami, N.~Gupta and D.~P.~Roy,
Phys.\ Rev.\ D {\bf 64}, 053002 (2001).

\bibitem{Choubey:2001ws}
S.~Choubey, S.~Goswami, K.~Kar, H.~M.~Antia and S.~M.~Chitre,
Phys.\ Rev.\ D {\bf 64} (2001) 113001

\bibitem{our-snocc}
A.~Bandyopadhyay, S.~Choubey, S.~Goswami and K.~Kar,
Phys.\ Lett.\ B {\bf 519}, 83 (2001). 

\bibitem{Choubey:2002nc}
S.~Choubey, A.~Bandyopadhyay, S.~Goswami and D.~P.~Roy,
arXiv:hep-ph/0209222.

\bibitem{our-snonc2}
A.~Bandyopadhyay, S.~Choubey, S.~Goswami and D.~P.~Roy,
Phys.\ Lett.\ B {\bf 540}, 14 (2002).

\bibitem{Bandyopadhyay:2002qg}
A.~Bandyopadhyay, S.~Choubey and S.~Goswami,
Phys.\ Lett.\ B {\bf 555}, 33 (2003).

\bibitem{others}
V.~Barger, D.~Marfatia, K.~Whisnant and B.~P.~Wood,
Phys.\ Lett.\ B {\bf 537}, 179 (2002);
%
P.~Creminelli, G.~Signorelli and A.~Strumia,
[arXiv:hep-ph/0102234 (version 3)];
%
J.~N.~Bahcall, M.~C.~Gonzalez-Garcia and C.~Pena-Garay,
JHEP {\bf 0207}, 054 (2002)
%
; 
P.~C.~de Holanda and A.~Yu.~Smirnov,
arXiv:hep-ph/0205241;
%
G.~L.~Fogli, E.~Lisi, A.~Marrone, D.~Montanino and A.~Palazzo,
arXiv:hep-ph/0206162.

\bibitem{Gonzalez-Garcia:2000dj}
M.~C.~Gonzalez-Garcia, C.~Pena-Garay and A.~Yu.~Smirnov,
Phys.\ Rev.\ D {\bf 63}, 113004 (2001).

\bibitem{kl}
K.~Eguchi {\it et al.}  [KamLAND Collaboration],
Phys.\ Rev.\ Lett.\  {\bf 90}, 021802 (2003).

%
\bibitem{Bahcall:2002ij}
J.~N.~Bahcall, M.~C.~Gonzalez-Garcia and C.~Pena-Garay,
arXiv:hep-ph/0212147.

\bibitem{raj}
A.~Bandyopadhyay, S.~Choubey, R.~Gandhi, S.~Goswami and D.~P.~Roy,
arXiv:hep-ph/0211266.

\bibitem{Bandyopadhyay:2002en}
A.~Bandyopadhyay, S.~Choubey, R.~Gandhi, S.~Goswami and D.~P.~Roy,
arXiv:hep-ph/0212146.

\bibitem{Fogli:2002au}
G.~L.~Fogli, E.~Lisi, A.~Marrone, D.~Montanino, A.~Palazzo and A.~M.~Rotunno,
arXiv:hep-ph/0212127;

\bibitem{deHolanda:2002iv}
P.~C.~de Holanda and A.~Yu.~Smirnov,
arXiv:hep-ph/0212270.
%
\bibitem{Bandyopadhyay:2003du}
A.~Bandyopadhyay, S.~Choubey and S.~Goswami,
arXiv:hep-ph/0302243, to appear in Phys. Rev.D. 

\bibitem{scho}The talk by S. Sch\"{o}nert,
http://neutrino2002.ph.tum.de.

\end{thebibliography}
\end{document}